\documentstyle[11pt,psfig]{article}
\def\i{\item}
\def\bi{\bibitem{}}

\def\ni{\noindent}
\def\beb{\begin{thebibliography}{999}}
\def\eeb{\end{thebibliography}}
\def\bei{\begin{itemize}}
\def\eei{\end{itemize}}
\def\bef{\begin{figure}}
\def\eef{\end{figure}}
\def\ben{\begin{enumerate}}
\def\een{\end{enumerate}}
\def\beq{\begin{equation}}
\def\eeq{\end{equation}}
\def\ber{\begin{eqnarray}}
\def\eer{\end{eqnarray}}

\newcommand{\dmdt}{{\mbox{{\rm M}$_{\odot}$}} {\rm yr}$^{-1}$}

\newcommand{\gcc}{{\rm g} \, {\rm cm}^{-3}}

\newcommand{\mdot}{\mbox{$\dot{M}$}}

\begin{document}
\centerline{\Large Evolution of the Magnetic Field in Accreting Neutron Stars}
\centerline{Sushan Konar} 
\centerline{\small {Raman Research Institute, Bangalore 560080, India}}
\centerline{\small {JAP, Indian Institute of Science, Bangalore 560012, India}}
\baselineskip=12pt

%\ni {\em Key Words} : magnetic fields, stars- neutron, pulsars-general, binaries-general \\

\section{Introduction}

\ni There has been sufficient observational indication suggesting a causal connection between the binary history 
of neutron stars and the evolution of their magnetic field. In particular, recent observations and their statistical
analyses suggest that (see Bhattacharya 1995, 1996, Bhattacharya \& Srinivasan 1995, Ruderman 1995 for detailed reviews):
\ben
\item Isolated pulsars with high magnetic fields ($ \sim 10^{11}$ -- $10^{13}$ G) do not undergo any significant 
field decay during their lifetimes;
\item Binary pulsars as well as millisecond and globular cluster pulsars, almost always having a binary history, 
possess much lower field strengths, down to $\sim 10^8$ G;
\item Most of these low-field pulsars are also quite old (age $\sim 10^9$ yrs), which implies that their field must 
be stable over such long periods; 
\i Binary pulsars with massive companions, like the Hulse-Taylor pulsar, have field strengths in excess of $10^{10}$~Gauss, 
whereas the low mass binary pulsars include both high-field pulsars and very low-field objects like the millisecond pulsars. 
\i The evolutionary link between millisecond pulsars and low-mass X-ray binaries seem to be borne out both from binary 
evolution models and from the comparative study of the kinematics of these two populations. 
\een

\ni Therefore, in this thesis we have tried to understand the mechanism of field evolution in neutron stars that are 
members of binary systems. To this end we have looked at four related problems as described below : 
\bei
\i the effect of diamagnetic screening on the final field of a neutron star accreting material from its
binary companion;
\i evolution of magnetic flux located in the crust of an accreting neutron star;
\i application of the above-mentioned model to real systems and a comparison with observations;
\i an investigation into the consequences of magnetic flux being initially located in the core of the star
and its observational implications.
\eei

\section{Effect of Diamagnetic Screening}
\ni We investigate the effect of diamagnetic screening on the final field of an accreting neutron star, in particular
we try to answer the following questions :
\ben
\i Are diffusive time-scales, in the layers where the field is expected to be buried, long enough to allow
screening to be an effective mechanism for long-term field reduction ?
\i Is it at all possible to bury the field or create horizontal components at the expense of the vertical
one against Rayleigh-Taylor overturn instability ?
\een
We find that,
\ben
\i The density of flow increases with an increase in the field strength. The larger the field, the deeper
and denser it gets buried . The flow density also has a mild positive dependence on the rate of mass accretion. 
But even for very large values of the surface field strength the flow does not occur at densities beyond 
$\sim 10^9~\gcc$. That means the flow always takes place within the liquid layer. And the earlier contentions of 
a burial within the solid layer does not really happen.
\i As the screening time-scales are always much smaller than the diffusive time-scales a condition of flux-freezing 
prevails and material movement indeed should proceed dragging the field lines along.
\i But, since the instability time-scale (sum of the overturn and the re-connection time-scales) is much too smaller 
than the other two time-scales of the problem, (fig. \ref{ftime-scale}) any stretching of field line is quickly 
restored (over the instability time-scale) and the material effectively flows past the field lines without causing 
any change.
\een
Therefore, it is not possible to screen the magnetic field of a neutron star by the accreting material in order to 
reduce the magnitude of the external dipole observed.

\bef
\begin{center}{\mbox{\psfig{file=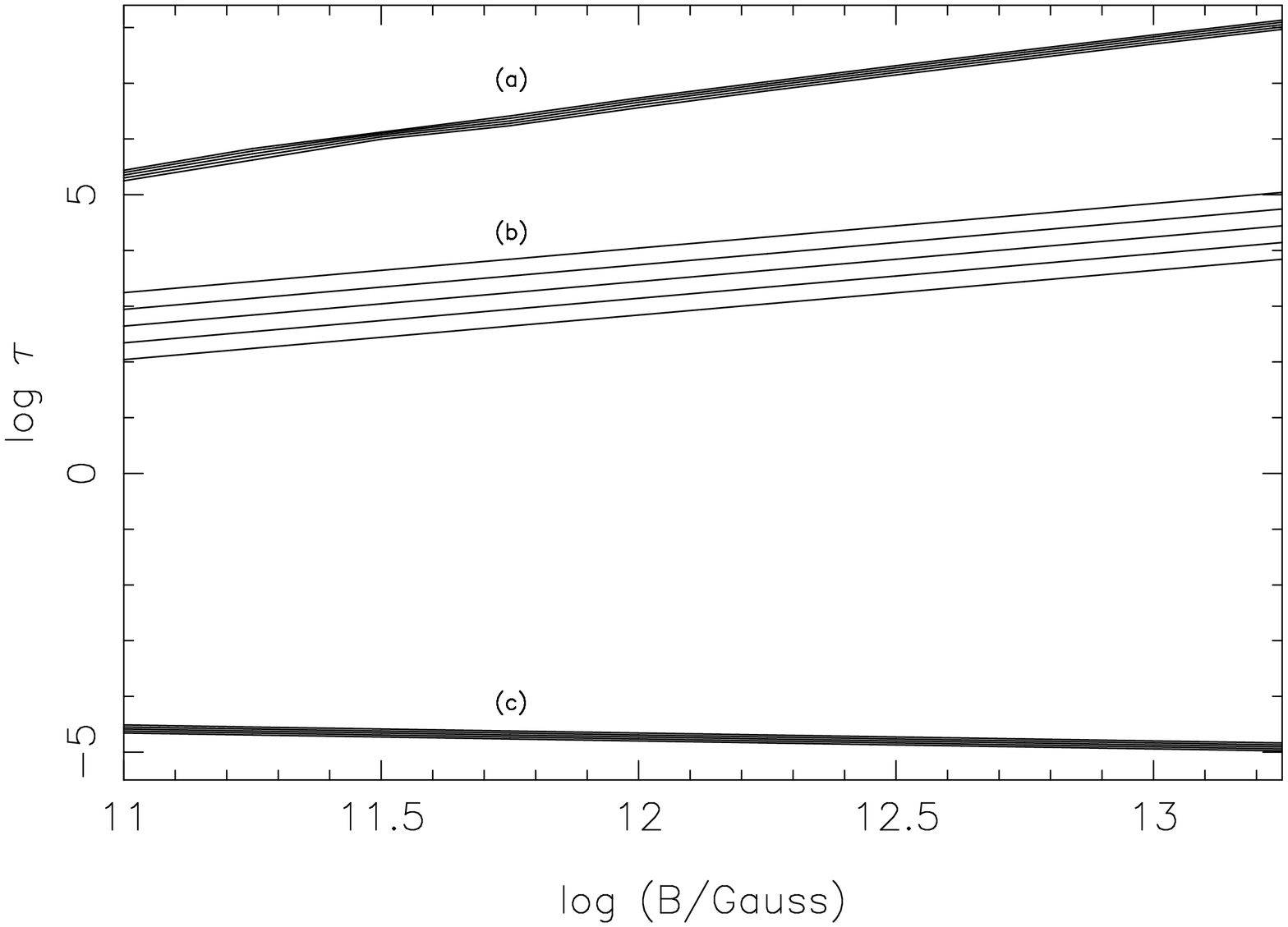,width=225pt,angle=-90}}}\end{center}
\caption[]{Variation of the time-scales with surface fields strength. The (a), (b), and (c) groups of curves
correspond to diffusive, screening and instability time-scales respectively. The curves (in each group)
correspond to $\mdot = 10^{-10}, 10^{-9.5}, 10^{-9}, 10^{-8.5}$~\dmdt respectively, the upper curves being of
lower values of accretion.}
\label{ftime-scale}
\eef

\section{Evolution of Crustal Magnetic Field in an Accreting Neutron Star}
\ni A possible mechanism of field decay is that of rapid ohmic diffusion in the accretion heated crust. The effect of 
accretion on purely crustal fields, for which the current loops are completely confined within the solid crust, is 
two-fold. On one hand the heating reduces the electrical conductivity and consequently the ohmic decay time-scale
inducing a faster decay of the field. At the same time the material movement, caused by the deposition of matter on 
top of the crust, pushes the original current carrying layers into deeper and denser regions where the higher conductivity 
slows the decay down. The mass of the crust of a neutron star changes very little with a change in the total mass; accretion 
therefore implies assimilation of the original crust into the superconducting core. When the original current carrying 
regions undergo such assimilation, further decay is stopped altogether. \\

\bef
\begin{center}{\mbox{\psfig{file=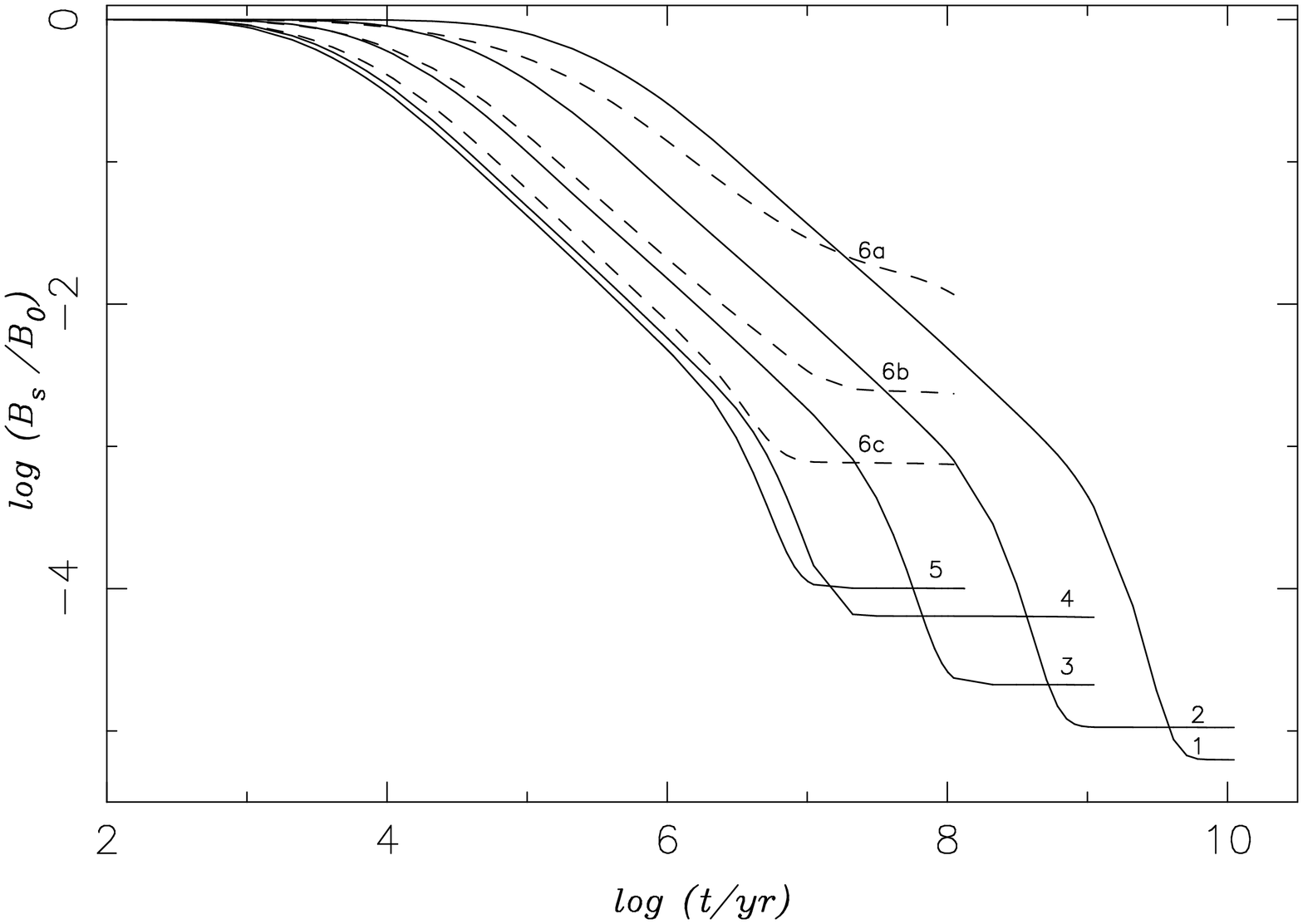,width=250pt,angle=-90}}}\end{center}
\caption[]{Evolution of the surface magnetic field for six values of accretion rate. The curves
1 to 5 correspond to $\mdot = 10^{-13}, 10^{-12}, 10^{-11}, 10^{-10}, 2.0 \times 10^{-10}$~\dmdt
with the crustal temperatures obtained from $\log T = 0.397 \log \mdot + 12.35$. The dashed curves 6a, 
6b and 6c correspond to $T = 10^{8.0}, 10^{8.25}, 10^{8.5}$~K respectively for an accretion rate of
$\mdot = 10^{-9}$~\dmdt. All curves correspond to $Q$ (impurity content of the crust) = 0.0, but are 
insensitive to the value of $Q$.}
\label{fb_evolv1}
\eef

\ni The combination of enhanced ohmic diffusion due to crustal heating and the transport of current-carrying layers to
higher densities due to the accreted overburden, causes the surface field strength to exhibit the following behaviour
(fig. \ref{fb_evolv1}):
\ben
\i An initial rapid decay (power law behaviour followed by exponential behaviour) followed by a leveling off (freezing),
\i Faster onset of freezing at higher crustal temperatures and at a lower final value of the surface field,
\i Lower final fields for lower rates of accretion for the same net amount of accretion,
\i The longer the duration of the pre-accretion phase the less the amount of field decay during the accretion phase, and
\i The deeper the initial current loops are the higher the final surface field.
\een
\ni For details see Konar \& Bhattacharya (1997).

\section{Comparison with Observation}
\ni We have investigated the evolution of the magnetic field of neutron stars in its entirety -- in case of the isolated 
pulsars as well as in different kinds of binary systems, assuming the field to be originally confined in the crust. 
We model the full evolution of a neutron star in a binary system through several stages of interaction. Initially there 
is no interaction between the components of the binary and the evolution of the neutron star is similar to that of 
an isolated one. It then interacts with the stellar wind of the companion and finally a phase of heavy mass 
transfer ensues through Roche-lobe overflow.  \\

\bef
\begin{center}{\mbox{\psfig{file=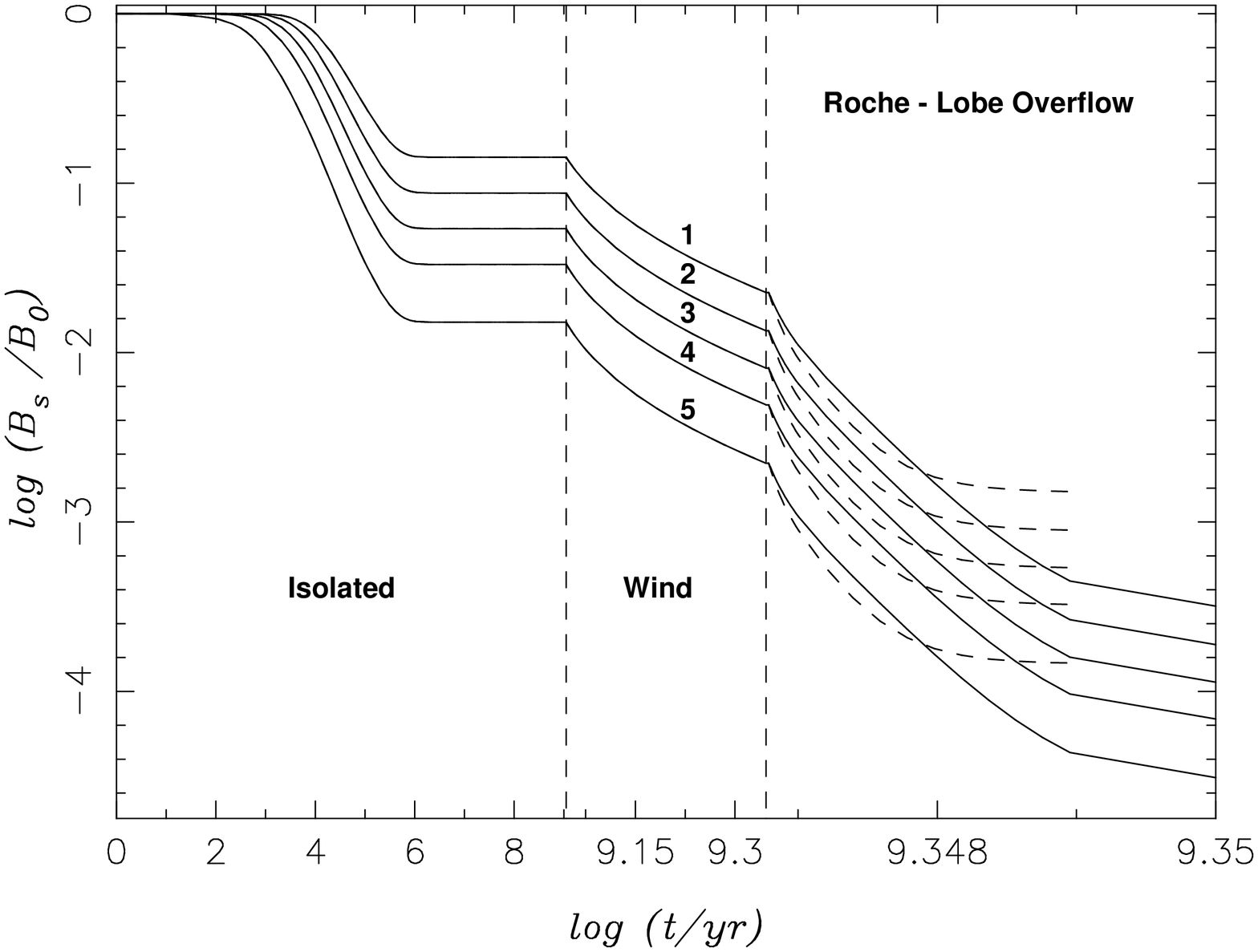,width=200pt,angle=-90}}}\end{center}
\caption[]{Evolution of the surface magnetic field in LMXBs with an wind accretion rate of $\mdot = 10^{-14}$~\dmdt. 
Curves 1 to 5 correspond to initial current configurations centred at 
$\rho = 10^{13}, 10^{12.5}, 10^{12}, 10^{11.5}, 10^{11} \gcc$. All curves correspond to $Q$ = 0.0. A standard cooling 
has been assumed for the isolated phase. The wind and Roche-contact phases are plotted in expanded scales. The dashed 
and the solid curves correspond to accretion rates of $\mdot = 10^{-9}, 10^{-10}$~\dmdt in the Roche-contact phase.}
\label{flmxb}
\eef

\ni We find that the model can explain almost all the features that have been observed to date. And our conclusions can be 
summarized as follows :
\bei
\i for this model to be consistent with the statistical analyses performed on the isolated pulsars at the
most a maximum value of 0.05 for the impurity strength can be allowed;
\i HMXBs produce high-field long period pulsars provided the duration of the wind accretion phase is short
or the initial current distribution is located at higher densities;
\i Relatively low-field ($B \sim 10^{10}$~Gauss) objects near death-line (low-luminosity pulsars) are also
predicted from HMXBs;
\i LMXBs will produce both high-field long period pulsars as well as low-field short period pulsars inclusive
of millisecond pulsars in the later variety (fig. \ref{flmxb}); and
\i a positive correlation between the rate of accretion and the final field strength is indicated that is
supported by observational evidence.
\eei
\ni For details see Konar \& Bhattacharya (1999a).

\section{Spin-down induced Flux-expulsion and its Consequences}
\ni Finally, we look at the outcome of spindown-induced expulsion of magnetic flux originally confined to the core, 
in which case the expelled flux undergoes ohmic decay. The general nature of field evolution again seems to fit the 
overall observational scenario. The nature of field evolution is quite similar to that in case of a purely crustal 
model of field evolution though the details differ. Most significantly, this model has the requirement of large values 
of the impurity strength $Q$ in direct contrast to the crustal model. To summarize then :
\bei
\i The field in isolated neutron stars do not undergo any significant decay, over the active lifetime of the pulsar,
conforming with the statistical analyses.
\i The field values in the high mass X-ray binaries can reaming fairly large for a moderate range of impurity strength.
\i A reduction of three to four orders of magnitude in the field strength can be achieved in the low mass X-ray binaries
provided the impurity strength is as large as 0.5 (fig. \ref{fcore_lmxb}).
\i If the wind accretion phase is absent then to achieve millisecond pulsar field values, impurity strength in excess
of unity is required.
\eei

\bef
\begin{center}{\mbox{\psfig{file=core_lmxb.ps,width=75pt,angle=-90}}}\end{center}
\caption[]{Evolution of the surface magnetic field in low mass X-ray binaries. The dotted and the solid curves correspond
to accretion rates of $\mdot = 10^{-9}, 10^{-10}$~\dmdt in the Roche contact phase.  The curves 1 to 5 correspond to
$Q$ = 0.0, 0.01, 0.02, 0.03 and 0.04 respectively. All curves correspond to a wind accretion rate of $\mdot = 10^{-16}$~\dmdt. }
\label{fcore_lmxb}
\eef

\ni For details see Konar \& Bhattacharya (1999b).

\section{Conclusions}

\ni In this thesis we have mainly investigated two models of field evolution - that of an initial crustal current 
supporting the field and that of spin-down induced flux expulsion, also looked at the effect of diamagnetic 
screening in an accreting neutron star.  Following are our conclusions regarding the nature of field evolution 
in the crust.
\bei
\i Pure Ohmic Decay in Isolated Neutron Stars :
\ben
\i A slow cooling of the star gives rise to a fast decay and consequent low final field. The opposite 
happens in case of an accelerated cooling.
\i An initial crustal current distribution concentrated at lower densities again gives rise to faster decay 
and low final surface field. Whereas if the current is located at higher densities the decay is slow resulting
in a higher final surface field.
\i A large value of impurity strength implies a rapid decay and low final field. If the crust behaves more like
a pure crystal the decay slows down considerably. 
\een
\i Accretion-Induced Field Decay in Accreting Neutron Stars :
\ben
\i In an accreting neutron star the field undergoes an initial rapid decay, followed by slow down and an eventual 
{\em freezing}. 
\i A positive correlation between the rate of accretion and the final field strength is observed, giving rise to
higher final saturation field strengths for higher rates of accretion.
\i An expected screening of the surface field by the diamagnetic accreting material is rendered ineffective by the 
interchange instabilities in the liquid surface layers of the star.
\i To produce millisecond pulsars in LMXBs in spin-down induced flux expulsion model very large values of impurities 
are required. This makes {\em the surface field go down to very low values in $10^9$~years in isolated pulsars} in 
contrast to a purely crustal model.
\een
\eei

\section*{Acknowledgement}
I would like to thank Dipankar Bhattacharya for his guidance and support.

\beb
\bi Bhattacharya D., 1995, JA\&A, 16, 227
\bi Bhattacharya D., 1996, {\em Pulsar Timing, General relativity and
the Internal Structure of Neutron Stars}, Royal Netherlands Academy of Arts and Sciences, in press.
\bi Konar S., Bhattacharya D., 1997, MNRAS, 284, 311
\bi Konar S., Bhattacharya D., 1999a, MNRAS, 303, 588
\bi Konar S., Bhattacharya D., 1999b, MNRAS, (in press)
\bi Ruderman M., 1995, JA\&A, 16, 207
\bi Bhattacharya D., Srinivasan G., 1995, {\em X-Ray Binaries}, ed.
Lewin W.~H.~G., van Paradijs J., van den Heuvel E.~P.~J., Cambridge University Press
\eeb

\end{document}